\documentclass[twocolumn,showpacs,preprintnumbers,amsmath,amssymb]{revtex4}
\usepackage{epsfig}

\begin{document}
\draft

\title{
Subharmonics and Aperiodicity in Hysteresis Loops
}

\author{Joshua M. Deutsch and Onuttom Narayan}
\affiliation{
Department of Physics, University of California, Santa Cruz, CA 95064.}

\date{\today}

\begin{abstract}
We show that it is possible to have  hysteretic behavior for magnets
that does not form simple closed loops in steady state, but must cycle
multiple times before returning to its initial state.  We show this by
studying the zero-temperature dynamics of the 3d Edwards Anderson spin
glass. The specific multiple varies from system to system and is often
quite large and increases with system size.  The last result suggests
that the magnetization could be aperiodic in the large system limit for
some realizations of randomness.  It should be possible to observe this
phenomena in low-temperature experiments.
\end{abstract}


\maketitle

Hysteresis in magnetic systems\cite{masters} is the prototype
for understanding history dependent behavior in all complex
systems.\cite{colloids,cdw,alloys} Although the usual example of this
is for ferromagnets, it can exist in other magnetic systems such as
anti-ferromagnets\cite{afm} or spin glasses.\cite{Zimanyi,Helmut}

For magnetic systems, when one cycles adiabatically between a {\em large}
positive and negative field $H$, the magnetization $M$ forms a closed loop
in the $H-M$ plane because of saturation. There is an implicit expectation
that even if the field were cycled repeatedly between two {\em moderate}
values, the magnetization would settle down to a steady state closed
loop, after some initial transient response. In this letter, we show
that this expectation is often {\em wrong}. Many cycles of the external
field are necessary to bring the magnetization back to its initial value,
and that as the system size grows, so does the number of cycles needed.

We do this for Ising spin glass models: both short range and long range
models violate the usual assumption of simple closed loops. Even in
steady state, the magnetization returns to its initial value only after
multiple cycles of the field. Another way of looking at this phenomenon,
probably more relevant to real experiments, is that with a steady state
oscillatory field the magnetization shows a subharmonic component. The
number of cycles it takes for the magnetization to return varies from
system to system, depending on the realization of the randomness, and can
be very large. Our estimates suggest that this effect should be observable
in nano-particles of spin glass material at very low temperatures.

This result might appear to be rather surprising as there are many cases
when the strong result of ``return point memory''\cite{Sethna,RPM} can
be shown: if a system starts at a field $H_1$ and ends at a field $H_2$,
its final state is independent of the time dependence of H as long as H
is never outside the interval $[H_1,H_2]$. In the case above this would
imply at most one transient cycle before the steady state closed loop was
reached.  An example of such a system is a random field Ising ferromagnet
where an elegant proof of return point memory was given.\cite{Sethna}
However the proof does not apply to cases where antiferromagnetic
interactions are also present, such as a spin glass. We show below that
in this case there is a much richer class of memory effects.

Earlier work has reported subharmonic response in the magnetization
of ferromagnetic materials subjected to oscillating magnetic
fields. \cite{engineers} It is a specific example of the general
phenomenon of ferroresonance.  However, this occurs when the magnetic
material is part of an electrical circuit (e.g. as the core of an
inductor) that is driven at finite frequency. Furthermore the intrinsic
response of the ferromagnetic material is expected to show return
point memory. From a dynamical systems viewpoint, this is simply the
observation that a driven nonlinear oscillator can respond at a frequency
different from the driving frequency, possibly depending on the initial
conditions of the system.  There has also been theoretical work on finite
frequency hysteresis loops in ferromagnets.\cite{Nattermann,Dhar1,Sides}
In contrast, we only consider $H$ fields that vary sufficiently slowly to
be adiabatic and our non-return point effect is intrinsic to the material.

We consider the standard Edwards-Anderson spin glass
Hamiltonian\cite{sg1,sg2} with Ising spins and nearest neighbor
interactions in three dimensions,
\begin{equation}
{\cal H} = -\sum_{\langle i,j\rangle} J_{i,j} S_i S_j - H\sum_i S_i .
\end{equation}
The coupling strengths $J_{i,j}$ are chosen to be uniform random numbers
between $-1$ and $1$. The Ising spin variables $S_i$ take values $\pm
1$. Because we expect to apply this model to granular materials, we
employ free boundary conditions. Other boundary conditions have been
verified to yield similar results.  We use single spin-flip dynamics at
zero temperature, and vary the magnetic field $H$ adiabatically. Thus as
the magnetic field is changed, a spin only flips if it is energetically
favorable for it to do so. Once a spin flips, it can render other spins
unstable.  If more than one spin is unstable, it is important to have
an algorithm that chooses the order that spins flip in a deterministic
manner, to avoid introducing randomness into the system through the
dynamics.  Accordingly we flip the spin that lowers the energy the
most. This process is repeated until all the spins are stable, after
which the magnetic field can be further varied.\cite{foot:dyn}

We start the system at large $H$ with all the spins positive and lower
$H$ to $H_{min}$ adiabatically. The field is then raised to $H_{max}$.
Thereafter, the field is cycled repeatedly between these two extremal
values. After an initial transient period, the system reaches steady
state.  In this steady state we determine the number of cycles of $H$
that are required for the system to repeat its configuration. This varies
randomly, depending on the coupling strengths $J_{i,j}$.

Clearly if $H_{min}$ and $H_max$ are too large in magnitude, the
magnetization will saturate at the extremal points of the $H$ cycles and
the behavior seen is trivial. On the other hand, if the range of $H$
is too small, hysteretic effects are minimal and behavior of the kind
above will be hard to observe.  By trying various values of the extremal
fields, we find that it is best to choose them to be about half of the
saturation field. Consequently we work with $H_{max}  = -H_{min} = -1.4$.

\begin{figure} 
\centerline{\epsfxsize=\columnwidth \epsfbox{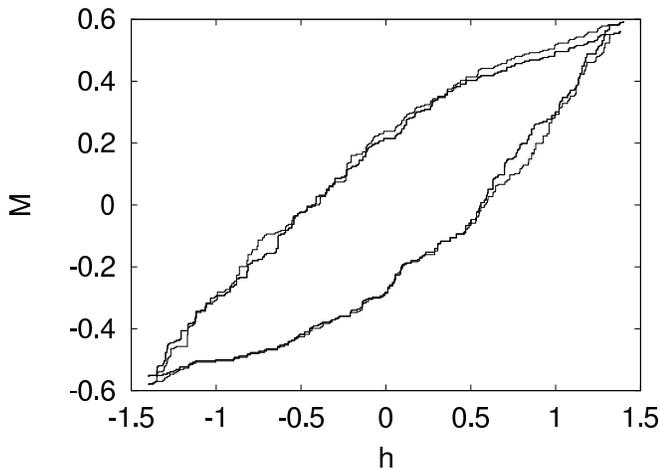}}
\caption{
Steady-state magnetization as a function of magnetic field for a spin-glass
with $8^3$ spins. It can be seen that two cycles of the magnetic field are
needed for the magnetization (and in fact the full spin configuration of the 
system) to return to its initial value.
}
\label{fig:loop}
\end{figure}

Figure \ref{fig:loop} shows an example of a system with linear dimension
$L=8,$ {\it i.e.\/} with $8^3$ spins, where the steady state behavior is a
two-cycle, {\it i.e.\/} two cycles of the magnetic field correspond to one
cycle in the magnetization. Even though the gap between the two halves
of the magnetization loop is much smaller than the width of the loop
itself, with an oscillatory magnetic field, it would be straightforward
to detect the subharmonic response.  Although more complicated cycles
are often found, they are not shown in the figure for the sake of clarity.

More generally, $m$ cycles of the magnetic field can be required
for the (steady state) magnetization to return to its initial value.
The distribution of $m$ for systems of different sizes is shown in
Table \ref{loop_length}. One observes that the larger the system size,
the more likely it is to find long cycles.  In fact for our largest
size, with $16^3$ spins, for about 1\% of the systems we were unable to
detect any periodicity even when checking for $m$ up to 128.  The fact
that for the remaining systems, where periodicity was seen, the biggest
value of $m$ detected was 42, suggests the existence of truly aperiodic
behavior\cite{periodicity_footnote}.

\begin{table}
\begin{center}
\begin{tabular}{|l|c|c|c|}
\hline
$\,\, m$&  L=4  &  L=8 & L=16 \\
\hline
$\,\,$1    &$\,\,$ 0.9343$\,\,$ & $\,\,$ 0.68 $\,\,$   & $\,\,$ 0.052 $\,\,$\\
$\,\,$2    & 0.0241  &  0.2342    & 0.503  \\
$\,\,$3    & 0.0359  &  0.0481    & 0.044  \\
$\,\,$4    & 0.0005  &  0.0179    & 0.109  \\
$\,\,$5    & 0.0044  &  0.0056    & 0.006  \\
$\,\,$6    & 0       &  0.0086    & 0.206  \\
$\,\,$7    & 0.0006  &  0.0021    & 0      \\
$\,\,$8    & 0       &  0.0010    & 0.007  \\
$\,\,$9    & 0.0002  &  0.0011    & 0.002  \\
$\,\,$10   & 0       &  0.0005    & 0.025  \\
$\,\,>$ 10 & 0       &  0.0009    & 0.046  \\

\hline
\end{tabular}
\end{center}
\caption{
Probability that $m$ cycles of the magnetic field are required for
the magnetization to return to its initial value, for systems of sizes
$4^3$, $8^3$ and $16^3.$ The number of systems considered for the three
different sizes was x, y and z respectively.  As $L$ is increased,
there is a greater probability of larger values of $m.$ }
\label{loop_length}
\end{table}

Although Table \ref{loop_length} shows an increase in typical cycle size
as the system is made larger, to some extent this is not caused by the
dynamics becoming more complex, but because of the way that the cycle
length is defined. For instance if the system were divisible into two
independent halves, with a 6 and 7 cycle respectively, the full system
would need 42 cycles to return to its original configuration. As the
system is made larger, the number of such subsystems could proliferate,
driving up the apparent cycle length. To eliminate this kind of
behavior, it is useful to examine the power spectrum of these systems.
One measures the magnetization at $H_{min}$ and $H_{max}$, the extrema
of the cycles of $H$. If the magnetization goes through an $m$-cycle,
these measurements will have a periodicity of $2m,$ while corresponding
measurements of $H$ would oscillate between $H_{min}$ and $H_{max}$,
exhibiting a periodicity of 2. The discrete Fourier transform of the
magnetization measurements is taken and the power spectrum is calculated.
In the case above, the spectrum will show peaks only at multiples of $f=
1/6$ and $f =1/7$, where $f=1$ is the frequency of the $H$ field, whereas
if the system were truly indivisible, peaks at multiples of $f=1/42$
would be observed. In addition, the power spectrum is crucial to detecting
these effects experimentally, as we have discussed in connection with
figure \ref{fig:loop}. Although in an experimental situation, one would
measure the power spectrum for the full time dependent magnetization
$M(t)$ instead of considering only its values at the extrema of $H$,
the discrete Fourier transform computed here should be a good indicator
of the experimental results.

\begin{figure} 
\centerline{\epsfxsize=\columnwidth \epsfbox{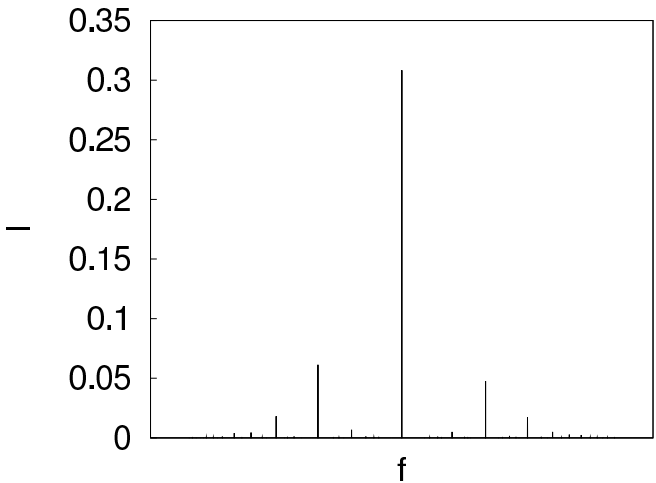}}
\caption{
Average power spectrum for systems of $8^3$ spins. The peaks at $f=1$
and $f=0$, whose heights are approximately $3\times 10^4$, have been
suppressed. The power is concentrated at $f=1/2,$ with smaller peaks at
other fractions.}
\label{fig:pow}
\end{figure}

\begin{figure}
\centerline{\epsfxsize=\columnwidth \epsfbox{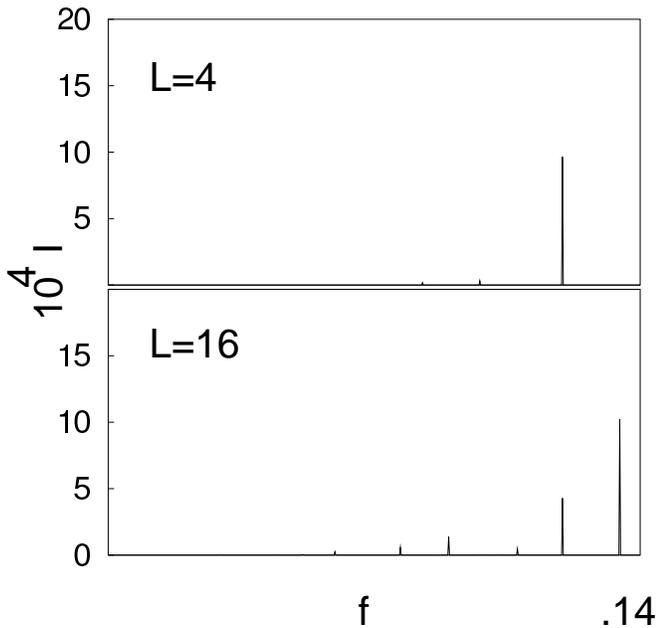}}
\caption{
Low frequency average power spectrum for systems of $4^3$ and $16^3$
spins. (To avoid confusion with peaks, no tick marks are shown on 
the $x$ axis.) There is substantially more power at low frequencies
for $16^3$ spins, and the spectrum extends down to $1/f\approx 40,$
{\it i.e.\/} a 40-cycle that is indivisible.}
\label{fig:lowfreq}
\end{figure}

Figure \ref{fig:pow} shows the power spectrum for $L=16$.  This is
an average of the spectra for all the different $m$-cycles, weighted
according to their probabilities (given in Table~\ref{loop_length}.
Sharp peaks are seen at discrete frequencies, with the dominant peak
at $f=1/2$. Figure \ref{fig:lowfreq} is an enlarged plot of the power
spectrum at low frequencies for $L=4$ and $L=16.$ Although small, the
low frequency power is clearly greater for $L=16.$ In addition, the 
peaks go down to lower frequencies, demonstrating that the dynamics
get more complex as $L$ is increased in addition to the trivial 
growth in cycle length discussed in the previous paragraph.

In Figure \ref{fig:pow} we have not plotted the two trivial peaks, at
$f=0$ and $1$.  These are approximately $10^5$ times the strength of the
peak at $f=1/2$.  As one would expect from Table.~\ref{loop_length}, the
corresponding power spectrum for $L=4$ has much fewer peaks. However, the
strength of the peaks is larger: the peak at $f=1/2$ is approximately
$10^{-4}$ times the strength of the peaks at $f=0$ and $1.$ More
generally, as $L$ is varied, the strength of the trivial peaks will be
proportional to the number of spins, i.e. $L^3$. Although the limited
range of $L$ that we have been able to simulate does not allow us to
determine the dependence on $L$ of the other peaks, for large $L$ we might
expect any system to behave as a collection of reasonably independent
regions, and the strength of the low frequency peaks to be proportional
to $L^{3/2}$. However, this argument is hazardous, since even a small
coupling between different regions of the system might affect subtle
details of the dynamics.  On the basis of our numerics, we make the more
conservative claim, that the strength of the low frequency peaks increases
with $L$, while their relative strength (compared to the peaks at $f=0$
and 1) decreases with $L.$ In a phase-locked experimental setup, the
absolute rather than the relative strength of these peaks is important.

We now turn to experimental considerations. Nanoparticles of spin glass
material, such as CuMn, with of order $16^3$ separate spins would have a
linear dimension of a few nanometers, which could be readily manufactured.
In fact it should also be possible to reduce the grain size somewhat,
although as discussed in the previous paragraph the subharmonic content
is reduced. In order to eliminate possible problems from interaction
between the grains, it would advisable to coat or embed the grains in a
neutral material to isolate them sufficiently from each other. With $N_g$
independent grains, the strength of the signal would be proportional
to $\sqrt{N_g}.$

As discussed earlier, the field should be varied over a range of
order half the saturation field, which would be $\sim 1T$.  The time
dependence of the magnetic field should be sufficiently slow so as to
evolve the spins adiabatically, and to prevent any significant eddy
current heating. For the small grains that we consider, a frequency of
the order of $10 Hz$ should be adequate.

The critical temperature $T_c$ of $CuMn$ is of order $1K$; in order to
see the effects that we have discussed, which are very small compared
to the overall hysteresis loop, it is necessary to essentially eliminate
thermal noise over the time scales we are interested in.  With activated
dynamics, it should be possible to achieve this if  the temperature is
of order $10 mK$.

It should be possible to satisfy all these conditions in an experiment
without too much difficulty. We note that recent work has found super spin
glass behavior in nanoparticle multilayer composites~\cite{ssg} and in
frozen ferrofluids~\cite{nordblad}, where the effective individual spins
are large and  $T_c$ is correspondingly high. While this would render
the phenomena we wish to observe accessible at higher temperatures,
one should be careful because the saturation field increases with $T_c$.

From a theoretical perspective, it is of interest to ask whether the
results we have found here are generic for systems with hysteresis. As
mentioned earlier in this paper, there are strong constraints on
many ferromagnetic models that force them to have return point
memory, precluding the possibility of multi-cycle hysteresis loops.
Even with the spin glass model that we have considered in this
paper, we find numerically that in one dimensional systems, when $H$
is cycled repeatedly, the hysteresis loop always settles down to a
one-cycle. However, steady state multi-cycles are seen in two dimensions,
at rates comparable to those for three dimensional systems.

From a dynamical systems approach, it would be natural to expect a system
with many interacting degrees of freedom to exhibit complex dynamical
behavior, possibly even chaos. In order to verify whether this is indeed
the source of the results reported here, we have carried out similar
numerical simulations on the Sherrington Kirkpatrick model~\cite{SK} for
spin glasses, as an example of a coupled non-linear system with several
degrees of freedom (the spins). We find that one needs a minimum of $5$
spins to find multi-cycle steady state magnetization.  In this case,
one finds only three cycles and no others aside from the trivial closed
loop case. The fraction of systems showing three cycles for $5$ spins is
low, approximately $3\times10^{-4}$. However this fraction rises steeply
with the number of spins.  One has to be careful in applying general
results from dynamical systems, since the dynamics we consider here are
of a special kind, being driven by energy minimization. For instance,
with a time-independent magnetic field and any initial configuration,
the system must evolve towards a fixed point. Further work in this
direction may explain the essential ingredients necessary to see the
behavior reported in this paper.

Our results for the SK model suggest that systems which are described
by a Landau theory with five or more components to the order parameter
might, in the correct parameter regime, show multi-cycle hysteresis
behavior. This would have the advantage that the effect would show up
in intensive quantitites. As discussed earlier in this paper, for the
spin-glass system we have considered here the non-return point effects
for the average magnetization per spin should vanish in the large size
limit.~\cite{Zimanyi}

In conclusion, we have shown in this paper that hysteresis loops have
much richer behavior than commonly assumed. For Ising spin glass models
with zero temperature dynamics, a slowly varying cyclical magnetic field
produces a subharmonic response in the magnetization. Our estimates
indicate that this effect should be observable in low temperature
experiments. Broader implications for other systems have been discussed.

We thank David Belanger, Peter Young and Steve Ford for useful discussions.


\begin{references}
\bibitem{foot:dyn} The dynamics are the same as those of Ref.\cite{Helmut},
but in that paper it is not specified what procedure is adopted when
several spins become unstable.
\bibitem{periodicity_footnote}Although
for any finite system it is inevitable that it will eventually
become periodic.
\end{references}
\end{document}